\documentclass[12pt]{article}
\usepackage{latexsym,epsfig,amssymb,euscript,amsmath,verbatim}

\newcommand{\nn}{\nonumber}

\newcommand{\stau}{\tilde\tau}

\newcommand{\misse}{\slash\!\!\!\!E}
\newcommand{\misset}{\slash\!\!\!\!E_T}

\begin{document}

\pagestyle{empty}
\rightline{KEK: TH-1514}

\begin{center}

{\LARGE{\bf Collider signatures of goldstini\\
\medskip
in gauge mediation}}

\vspace{0.6cm}

{\large{Riccardo Argurio$^{1,5}$, Karen De Causmaecker$^{2,5}$, Gabriele Ferretti$^{3}$ \vskip 3pt  Alberto Mariotti$^{2,5}$, Kentarou Mawatari$^{2,5}$, and Yoshitaro Takaesu$^{4}$  \\[5mm]}}

{\small{{}$^1$ Physique Th\'eorique et Math\'ematique\\
Universit\'e Libre de Bruxelles, C.P. 231, 1050 Bruxelles, Belgium\\
\medskip
{}$^2$ Theoretische Natuurkunde and IIHE/ELEM\\
Vrije Universiteit Brussel, Pleinlaan 2, 1050 Brussels, Belgium\\
\medskip
{}$^3$  Department of Fundamental Physics \\
Chalmers University of Technology, 412 96 G\"oteborg, Sweden\\
\medskip
$^4$  KEK Theory Center, and Sokendai, Tsukuba 305-0801, Japan\\
\medskip
$^5$ International Solvay Institutes, Brussels, Belgium}}

\vspace{0.7cm}

{\bf Abstract}

\vspace{0.4cm}

\begin{minipage}[h]{16.0cm}

We investigate the collider signatures of the multiple goldstini
scenario in the framework of gauge mediation. This class of models is
characterized by a visible sector (e.g.~the MSSM or any extension)
coupled by gauge
interactions to more than one SUSY breaking sector. The spectrum
consists of a light gravitino LSP, behaving as a goldstino, and a
number of neutral fermions (the pseudo-goldstini) with a mass between
that of the LSP and that of the lightest particle of the observable
sector (LOSP). We consider the two 
situations where the LOSP is either a gaugino-like
neutralino or a stau and we assume only one pseudo-goldstino of a mass of
${\cal O}(100)$~GeV. The coupling of the
LOSP to the pseudo-goldstino can be enhanced with respect to those of
the gravitino giving rise to characteristic signatures. We show that
the decay modes of the LOSP into a SM particle and a
pseudo-goldstino can be significant.
For both LOSP scenarios we analyze (pseudo)-goldstini
production at colliders.
Compared to standard gauge mediation the final state spectrum is softer and
more structured.

\end{minipage}

\end{center}

\newpage

\setcounter{page}{1} \pagestyle{plain} \renewcommand{\thefootnote}{\arabic{footnote}} \setcounter{footnote}{0}

\section{Introduction}

In the coming years we will receive the final verdict from the
LHC experiments as to whether low energy supersymmetry (SUSY)
is present or not in nature. So far, both ATLAS and CMS experiments have
not seen any signal and imposed strong bounds on the masses of colored
superpartners~\cite{Chatrchyan:2011zy,Aad:2011ib,Collaboration:2011iu}.

In this paper we will work in the general framework of gauge mediation, where the SUSY breaking effects are communicated to the observable sector by gauge interactions and gravity effects are subleading. A subclass of these models, namely those with a light neutralino and a low SUSY breaking scale, has already been constrained by the search for prompt photons plus missing
energy~\cite{arXiv:0910.3606,Chatrchyan:2011wc,Chatrchyan:2011ah,Aad:2011kz,arXiv:1111.4116}.
Given that the most straightforward searches have not yielded results, it
is important at this stage to consider broadening the class of models to
assure that we are not missing anything.

In all generality, in a SUSY model based on gauge mediation the Lightest SUSY
Particle (LSP) is the gravitino (we consider models preserving R-parity).
The characteristic signatures of the model are then driven by the nature of the
Lightest Observable-sector SUSY Particle (LOSP) and by its decay properties.
The two most common scenarios are those with a neutralino or a stau LOSP.

Even within this general framework, it is important to
keep exploring less conventional avenues to make sure that
we get the most out of the current searches. One of the motivations for our work is to
explore the theoretical aspects and experimental signatures of the goldstini
scenario~\cite{Cheung:2010mc,Cheung:2010qf} (see also
\cite{Benakli:2007zza} for previous work and \cite{Craig:2010yf,McCullough:2010wf,Cheng:2010mw,Izawa:2011hi,Thaler:2011me,Cheung:2011jq,Bertolini:2011tw}
for follow up work), in the context of gauge mediation as previously considered
in~\cite{Argurio:2011hs}.

This scenario arises when there are more than one SUSY breaking hidden sectors interacting with the observable sector via gauge interactions.
Each sector gives rise to a goldstino, a linear combination of which is eaten by the gravitino while the remaining fields give rise to light, neutral, spin one-half particles, the pseudo-goldstini. For simplicity, in the following we are going to consider a set up
where there are only two decoupled hidden sectors and hence only one
pseudo-goldstino, but the analysis is trivially extended to the more general case and the qualitative features of the model are unchanged.

It was shown in~\cite{Argurio:2011hs} that the pseudo-goldstino
acquires a mass in the range 1--100 GeV from radiative corrections, while all supergravity effects can be neglected.
If this mass is lower than that of the LOSP, the pseudo-goldstino becomes the Next-to-Lightest SUSY
Particle (NLSP).

In this case one may face situations where the SUSY breaking scale is
low, and thus the gravitino is an almost massless LSP, behaving as
the massless true goldstino (denoted by $G$) up to a very good level of accuracy. Yet, the
LOSP decays predominantly through other channels, namely those containing
a massive pseudo-goldstino (denoted by $G'$). One of the consequences of this fact is
that it softens the spectra of the final states, and 
we investigate those effects for the two most likely situations 
where the LOSP is either a gaugino-like
neutralino or a stau. 

In the case of a neutralino LOSP (denoted by $\chi$) the most accessible signature arises when $\chi$ is mostly a gaugino, promptly decaying into a photon and a (pseudo)-goldstino. The final state will contain photons and missing energy.
A mostly gaugino $\chi$ can also decay into a $Z$ boson and we will
analyze these two situations in detail. It would also be interesting
to consider the case where the $\chi$ has a large higgsino component
enhancing its decay into a higgs.\footnote{Shorthand for
  Brout-Englert-Higgs scalar.} 
For a single sector this was considered in~\cite{arXiv:0911.4130}, (see also the recent work~\cite{Petersson:2012dp} where mono-photon signals arise directly for higgs decay).
In the case of the stau LOSP (denoted by $\stau$), the lepton number
conservation requires a stau to be created in pairs and the final state contains two taus and missing energy.



Our analysis is independent from many detailed features
of the spectrum of superpartners, in the spirit of
simplified models.  
When forced to pick specific values 
for sparticle masses and mixing angles
we will choose those used in common benchmark
points~\cite{Allanach:2002nj} such as SPS8 ($\chi$ LOSP) and SPS7
($\stau$ LOSP), in order to make a comparison with the standard gauge
mediated models. 
%
Other LOSP scenarios, e.g. higgsino-like neutralino LOSP or
colored LOSP suggested by general gauge mediation \cite{Meade:2008wd},
are also interesting 
to be applied within the goldstini framework and will be reported
elsewhere. 
 It should also be emphasized that although the points based on a minimal
GMSB model have been constrained by the Tevatron~\cite{Aaltonen:2009tp,Abazov:2010us} and by the
LHC~\cite{arXiv:1111.4116}, the multiple goldstini scenario eases the constraints due to the softer final spectrum.

We will  refrain from considering cosmological issues, just assuming a
long enough lifetime for the pseudo-goldstino such that it safely
escapes the detector (see e.g.~\cite{Cheng:2010mw,Argurio:2011hs} for
considerations on the lifetime appropriate to our set up).

The paper is organized as follows: Section~2 provides the theoretical background needed to describe the processes of interest.
We discuss the spectrum and the couplings of the pseudo-goldstini to the
Supersymmetric Standard Model (SSM). We mainly emphasize the differences from the well known couplings of the
true goldstino with SSM fields. As already mentioned above, in the many-goldstini scenario,
there are additional particles with masses in between the LOSP and the LSP, the pseudo-goldstini. They
come from other hidden sectors, but have couplings with SSM particles
dictated by their nature of being goldstini in each hidden sector.
The pseudo-goldstino has an interaction Lagrangian with the visible sector similar in structure
to the one of the true goldstino, with however couplings that are no
longer fixed in terms of the SSM parameters and the (overall) SUSY
breaking scale $F$, but are essentially additional parameters. They can be computed in a specific set up, but can also be
considered free as far as phenomenological models are concerned. They can easily be enhanced with respect to the goldstino
couplings. Given that this is the most original feature of these
models, we focus our interest on analyzing the dependence of the experimental signatures on these parameters and on the mass of the pseudo-goldstino for some appropriately chosen simplified spectrum for the observable sector.

In section~3 we discuss the case of neutralino LOSP.
We exemplify some of the features of these models by first
considering processes that do not involve the intricacies of strong
dynamics and can be treated analytically to a large extent. We first
study the possible decays of the neutralino into ($\gamma G$), ($\gamma G'$), ($ZG$) and ($ZG'$) pairs.
(As mentioned before, in this paper we restrict our attention to a mostly gaugino $\chi$, and do not consider its decays into a higgs.)
We then compute the production of neutralino and pseudo-goldstino in
electron-positron collision ($e^+e^-\to\chi G'$), which leads to the $\gamma+\misse$ signature.
The analytic result for the cross section has been used to test our FeynRules
implementation but it is also of interest in future
$e^+e^-$ linear colliders such as the ILC.
We then simulate the neutralino-pair production, $e^+e^-\to\chi\chi$,
which gives the $\gamma\gamma+\misse$ signal, also of interest for the ILC.

We compare how the total and differential cross sections change
in magnitude and shape with respect to the standard results obtained in a single goldstino scenario.
In particular, the total cross section of the single photon processes can be
enhanced considerably for a pseudo-goldstino since the couplings of
the latter are no longer strictly tied to the overall SUSY breaking scale.
Furthermore, the mass and the coupling of the pseudo-goldstino alters the shape
of the differential cross sections with respect to those of a nearly massless goldstino/gravitino,
giving rise to a softer and more structured spectrum for the photons.

We then proceed to the simulation for $pp$ collisions of relevance to the LHC.
We focus on the following exclusive processes without jets.
First we discuss the photon(s) plus missing energy signals.
As a more promising channel, the $l^+l^-+\gamma\gamma+\misset$ signature is studied in detail,
which is provided by pair production of sleptons,
subsequently decaying to a lepton and a neutralino LOSP.
While the total cross-sections are independent from the masses and
couplings of the pseudo-goldstino, the distributions of emitted photons
and missing transverse energy do
depend on these parameters and we illustrate this point by simulating
various examples.

In section~4 we discuss the case of stau LOSP. We first study the decay
widths and branching ratios of the stau to $(\tau G)$ and $(\tau G')$.  
We then consider stau pair production in $pp$ collisions, studying the 
$\tau^+\tau^-+\misset$ signature. 
There is large SM background, but a satisfactory significance can be
achieved by imposing some kinematical cuts.
We find again that the pseudo-goldstino alters the shape
of the differential cross sections with respect to the ordinary gauge
mediated scenario. 

We briefly conclude in section~5.

\section{The two-sector model and goldstini couplings}
\label{formal}
\subsection{General formalism} \label{generalform}

The purpose of this section is to derive the relevant couplings of the (pseudo)-goldstino.
We consider a set up where there are two completely decoupled hidden
sectors, each communicating to the SSM through gauge interactions. In
other words, each sector is a model of gauge mediation on its own.
(For a review of gauge mediation see~\cite{Giudice:1998bp}.) Each sector
will then have a goldstino, $G_1$ and $G_2$ respectively, coupling to
the SSM particles in a way dictated by the universal low-energy effective
action of the goldstino, applied to each sector. If $F_1$ and $F_2$
are the respective SUSY breaking scales of each sector, and
$F=\sqrt{F_1^2+F_2^2}$, then the true and pseudo-goldstino are given
by
\begin{align}
 G &= \frac{1}{F}(F_1 G_1 + F_2 G_2), \\
 G'&= \frac{1}{F}(-F_2 G_1 + F_1 G_2).
\end{align}
For definiteness, we will always assume $F_1>F_2$. It has been shown
in \cite{Argurio:2011hs} that, while $G$ can be considered massless to
all effects, $G'$ acquires radiatively a mass that is of the order of GeV if
the scales in the two hidden sectors are of the same order. Moreover,
when there is a hierarchy in the SUSY breaking scales $F_1\gg F_2$,
the mass is enhanced to
\begin{align}
 m_{G'} \sim \frac{F_1}{F_2}\, \mbox{GeV}.
\end{align}
Since the exact expression for $m_{G'}$ is very much model dependent, for phenomenological purposes
it should be taken as a free parameter, recalling that it will
reasonably fall in the 1--100 GeV range. We will always
assume that $G'$ is the NLSP, i.e.\ it is lighter than the LOSP
otherwise its presence is virtually impossible to detect.

The couplings of $G$ and $G'$ to the SSM particles can be derived by
considering the couplings of $G_1$ and $G_2$, each as if it were the
only source of SUSY breaking.

The physical processes we are interested in involve only vertices with
at most one (pseudo)-goldstino so we will not have to deal with the
intricacies of the full (pseudo)-goldstino Lagrangian. We start by
considering the couplings to a generic observable sector whose chiral
and vector multiplets we denote by $\Phi \ni (\phi, \psi_\alpha,
F_\phi)$ and ${\cal W_\alpha} \ni (\lambda_\alpha, A_\mu,  D)$
respectively (gauge and flavor indices suppressed). After discussing the
general features in this language, we present the explicit couplings in
the context of the SSM, which are slightly more involved due to the
additional presence of Electro-Weak Symmetry Breaking (EWSB). 

Let us begin by considering the true goldstino $G$, whose coupling at the linear level is fully understood
since the seminal work of the '70s~\cite{Fayet:1977vd}.
It is well known that there are two equivalent ways of writing the
linear coupling of $G$ to the matter fields. One is the derivative
coupling to the supercurrent
\begin{align}
   {\mathcal{L}}_{\partial} =  \frac{1}{F}(\partial_\mu G^\alpha J^\mu_\alpha + \mathrm{h.c.}),\label{derivativeaction}
\end{align}
where, in the conventions of~\cite{Dreiner:2008tw}
\begin{align}
   J^\mu = \sigma^\nu\bar\sigma^\mu \psi D_\nu\phi^* - i
   \sigma^\mu \bar \psi F_\phi +i \frac{1}{2\sqrt{2}}
            \sigma^\nu\bar\sigma^\rho\sigma^\mu \bar \lambda F_{\nu\rho} +
            \frac{1}{\sqrt{2}} \sigma^\mu\bar\lambda D.
\end{align}

The other action is the non-derivative coupling obtained from the previous one by integrating by parts and using the equation of motion to obtain $\Delta_\alpha = \partial_\mu J^\mu_\alpha|_{\mathrm{e.o.m.}}$
\begin{align}
   {\mathcal{L}}_\eth =  -\frac{1}{F}(G^\alpha \Delta_\alpha + \mathrm{h.c.}).\label{nonderivativeaction}
\end{align}
Since the non-conservation of the supercurrent is entirely due to the
presence of supersymmetry breaking soft terms, the expression for
$\Delta_\alpha$ must be a function of the latter.

Let us consider the most general soft SUSY
breaking terms, namely:
\begin{align}
    {\mathcal{L}}_{\hbox{\footnotesize soft}}=
       -\frac{1}{2} m_\lambda \lambda  \lambda - \frac{1}{2} m_\lambda^* \bar\lambda \bar\lambda - U(\phi, \phi^*),
\end{align}
where $U$ is a (at most cubic) gauge invariant function of the scalars, containing for
instance the sfermion soft masses and the $B\mu$ term and $m_\lambda$ are the Majorana masses for the gauginos.
The divergence of the supercurrent is now
\begin{align}
     \partial_\mu J^\mu_\alpha = \Delta_\alpha =
       \frac{m_\lambda}{2 \sqrt{2}} \sigma^\mu \bar\sigma^\nu
       \lambda_\alpha F_{\mu\nu} - i\frac{m_\lambda}{\sqrt{2}} \lambda_\alpha D -
       \psi_{\alpha} \frac{\partial U}{\partial \phi} \label{delta}
\end{align}
from which the coupling follows using (\ref{nonderivativeaction}).
This expression is valid regardless of whether the gauge symmetry is
spontaneously broken or not (see for instance~\cite{Luo:2010he} for a
complete treatment in the MSSM). The form of
(\ref{nonderivativeaction}) can also easily be derived from
a superspace formulation of the broken SUSY theory, in terms of the
goldstino superfield $X$.

The derivative and non-derivative actions are completely equivalent,
of course, but some issues are easier
to investigate in one formalism than in the other.

It turns out that it is the non-derivative action that is more directly
generalized to the pseudo-goldstino case.
This can be seen in various ways but perhaps the easiest argument
comes by looking at the leading high energy behavior of the $2\to 2$
scattering amplitudes involving a goldstino. The action
(\ref{derivativeaction}) contains terms of dimension six and one would
expect the tree-level unpolarized squared amplitudes to scale like
$s^2$ where $\sqrt{s}$ is the center of mass energy. This is not what
happens however since SUSY ensures that the leading order behavior
cancels between the different contributions. This must be so since
(\ref{derivativeaction}) is equivalent to (\ref{nonderivativeaction})
which contains terms of dimension at most five and yields a scaling of
order $s$.

The same scaling must occur for the pseudo-goldstino since, after all,
at tree level it is a linear combination of two decoupled
goldstini, but now the relative coefficients between the various terms
in the action are no longer fixed by SUSY. Using the non-derivative
action ensures that the high energy behavior is preserved. One can of
course use the equations of motion ``backwards'' and  rewrite the
non-derivative action in terms of the same type of terms that appear
in the derivative action but with different relative coefficients. In
the process however one also picks up dimension six contact terms
schematically like $G \lambda \psi \psi$ that once again cancel the
$s^2$ behavior. Thus it is clearly more convenient to work with
(\ref{nonderivativeaction}).

In order to derive the couplings of the pseudo-goldstino,
let us start at tree level with the two fields $G_h$ ($h=1,2$) that are the goldstini of the respective hidden sectors contained in the superfields $X_h$. For each goldstino-gaugino-gauge boson vertex, we have
\begin{align}
\frac{1}{2}\int d^2\theta \frac{m_{\lambda(h)}}{F_h}X_h {\cal W}^2
\supset \frac{ m_{\lambda(h)}}{2\sqrt{2}F_h}\lambda \sigma^\mu
\bar\sigma^\nu G_hF_{\mu\nu}. \label{GFL}
\end{align}
~For the goldstino-fermion-sfermion vertex, we have
\begin{align}
\int d^4\theta \frac{m^2_{\phi(h)}}{F_h^2}X_hX_h^\dagger \Phi\Phi^\dagger
\supset \frac{m^2_{\phi(h)}}{F_h}G_h\psi \phi^*. \label{Gsfs}
\end{align}
The terms (\ref{GFL}) and (\ref{Gsfs}) correspond to the first and last
terms in (\ref{delta}) inserted into (\ref{nonderivativeaction}), the
last one with $U = m_{\phi}^2 \phi^{*} \phi$. We ignore further vertices
that do not contribute to the processes of interest but a full treatment can be found in~\cite{Komargodski:2009rz}.

Rotating to the $G,G'$ basis, we obtain the couplings
\begin{align}
 \frac{m_{\lambda}}{2\sqrt{2}F}\lambda \sigma^\mu
\bar\sigma^\nu GF_{\mu\nu}+ K_\lambda \frac{
  m_{\lambda}}{2\sqrt{2}F}\lambda \sigma^\mu
\bar\sigma^\nu G'F_{\mu\nu}
\label{Glambdacoupl}
\end{align}
and
\begin{align}
\frac{m^2_{\phi}}{F}G\psi \phi^* +
K_\phi\frac{m^2_{\phi}}{F}G'\psi \phi^* ,
\label{Gphicoupl}
\end{align}
where $m_\lambda=m_{\lambda(1)}+m_{\lambda(2)}$ and
$m^2_{\phi}=m^2_{\phi(1)}+m^2_{\phi(2)}$, and
the factors $K_\lambda$ and $K_\phi$ are the ratios between the
coupling of the pseudo-goldstino to the one of the true goldstino.
Their expressions are
\begin{align}
K_\lambda &= -\frac{m_{\lambda(1)}}{m_\lambda}\frac{F_2}{F_1}
+\frac{m_{\lambda(2)}}{m_\lambda}\frac{F_1}{F_2}, \\
K_\phi &= -\frac{m_{\phi(1)}^2}{m_\phi^2}\frac{F_2}{F_1}
+\frac{m_{\phi(2)}^2}{m_\phi^2}\frac{F_1}{F_2},
\end{align}
where in general there will be different coefficients for each gauge group and matter multiplet.
In some specific models, there can be relations or bounds between
$K_\lambda$ and $K_\phi$. However, in the phenomenological set up, one assumes that these two parameters are free.
In any case, below we will use couplings of $G'$ only to a very limited set of
(s)particles. The interesting case will be when $K_\lambda,K_\phi\gg
1$. Note however that both cannot be larger than $F_1/F_2$, hence they
are limited to be somewhat smaller than $10^3$. (If $F_1/F_2$ is too
large, the above picture is no longer valid because the SUSY breaking
vacuum of the second sector is destabilized \cite{Argurio:2011hs}.)

\subsection{Pseudo-goldstino couplings in the SSM}

We now specialize to the case where the observable sector is a SUSY extension of the SM. A couple of technical issues arise since one must also rotate to the physical bases of mass eigenstates of the observable sector after EWSB. More generally the (pseudo)-goldstini will also mix with the particles in the observable sector carrying the same quantum numbers but it is possible to treat this mixing to first order in the SUSY breaking parameter.

Let us begin with the coupling of the neutralino to the $G/G'$ and the photon.
This involves  an element of the neutralino mixing matrix, essentially proportional
to how much the neutralino is the would-be photino. More precisely
for the goldstino there is a factor (see e.g.~\cite{Dreiner:2008tw})
\begin{align}
a_\gamma= N^*_{11}\cos \theta_W + N^*_{12} \sin\theta_W,
\end{align}
where $N^*_{11}$ and $N^*_{12}$ are the mixing angles between the
lightest neutralino and the Bino and Wino respectively. For the
pseudo-goldstino, we can multiply the above by a factor $K_{\gamma}$
that can be larger than one as discussed above.
The relevant part of the Lagrangian is thus
\begin{align}
{\cal L}_{\chi G\gamma} + {\cal L}_{\chi G'\gamma} =  \frac{a_\gamma m_\chi }{2\sqrt{2}F}\chi
\sigma^\mu \bar\sigma^\nu G  F_{\mu\nu} +  \frac{K_\gamma a_\gamma m_\chi }{2\sqrt{2}F}\chi
\sigma^\mu \bar\sigma^\nu G'  F_{\mu\nu}.  \label{chiphoton}
\end{align}

The coupling of the goldstino and the pseudo-goldstino to the
neutralino and the $Z$ boson is slightly more subtle. Standard
computations, such as in \cite{Ambrosanio:1996jn}, use the goldstino
couplings in derivative form. After EWSB, there are two such couplings:
\begin{align}
{\cal L}_{\partial} = i \frac{a_{Z_{T}}}{2\sqrt{2}F}\bar \chi \bar
\sigma^\mu \sigma^\nu \bar \sigma^\rho \partial_\mu G F_{\nu\rho}-
\frac{a_{Z_{L}}m_Z}{\sqrt{2}F} \chi \sigma^\mu \bar\sigma^\nu
\partial_\mu G Z_\nu +{\rm h.c.}, \label{derz}
\end{align}
where
\begin{align}
a_{Z_{T}}&= -N^*_{11}\sin \theta_W + N^*_{12} \cos\theta_W,\\
a_{Z_{L}}&= N^*_{13}\cos \beta - N^*_{14} \sin\beta.
\end{align}
$N_{13}^*$ and $N_{14}^*$ are the higgsino components of the lightest
neutralino, and $\tan\beta$ is the ratio of the vacuum expectation
values of the two higgs doublets\footnote{Although we do not consider the decay into higgs, we will nonetheless treat the decay into $Z$ exactly by including the contribution of the longitudinal component.}.
The notation follows from the notable fact that the first and
second term respectively
couple only transverse and longitudinal components of the $Z$ to a
massless goldstino.

Integrating by parts and using the equations of motion, one obtains
the following terms in the non-derivative Lagrangian:
\begin{align}
{\cal L}_{\chi G Z} =  \frac{a_{Z_{T}} m_\chi +a_{Z_{L}}m_Z}{2\sqrt{2}F}\chi
\sigma^\mu \bar\sigma^\nu G  F_{\mu\nu} +i
\frac{m_Z (a_{Z_{T}} m_Z +a_{Z_{L}}m_\chi)}{\sqrt{2}F}\bar\chi \bar
\sigma^\mu G Z_\mu + {\rm h.c.} \label{atal}
\end{align}
It might seem at first that the second term in the Lagrangian above
cannot be reproduced using (\ref{delta}). This term appears because
after EWSB there are off-diagonal mass terms involving the goldstino
and the neutralinos. The mass eigenstates have then to be shifted by
${\cal O}(1/F)$ terms mixing the goldstino with the
neutralinos. Eventually the term above arises from the neutralino
gauge couplings, and its precise form follows from using the equations
leading to the EWSB vacuum (for a discussion, also including the
pseudo-goldstino, see \cite{Thaler:2011me}).

The pseudo-goldstino couplings to the $Z$ boson will be of the form
above, with however some model dependent factors in front of each
term. Since there are two independent terms in (\ref{atal}), there will be two independent coefficients and we choose to associate them directly to the rescaling of the $a_{Z_{T}}$ and $a_{Z_{L}}$ coefficients:\footnote{One could have also chosen to rescale each Lorentz invariant term in (\ref{atal}) independently but this would obscure the comparison in the simulation.}
\begin{align}
{\cal L}_{\chi G' Z} &=  \frac{K_{Z_T}a_{Z_{T}} m_\chi +K_{Z_L}a_{Z_{L}}m_Z}{2\sqrt{2}F}\chi
\sigma^\mu \bar\sigma^\nu G'  F_{\mu\nu} \nn \\ &\quad +i
\frac{m_Z (K_{Z_T}a_{Z_{T}} m_Z +K_{Z_L}a_{Z_{L}}m_\chi)}{\sqrt{2}F}\bar\chi \bar
\sigma^\mu G' Z_\mu + {\rm h.c.} \label{ktkl}
\end{align}

We now move on to the couplings of the stau. Even in this case the
general formula (\ref{Gphicoupl}) requires some well known modifications
after EWSB. 
Namely, the presence of a SUSY contribution to the tau mass makes the
coefficient in the action depend on the difference of the masses in the
multiplet. The large soft off-diagonal corrections to the stau mass
matrix require rotating from the gauge eigenbasis $\stau_L, \stau_R$ to
the mass eigenbasis $\stau, \stau'$.
For the couplings to the first two families these effects can be neglected. We focus only on the LOSP 
$\stau =\cos\theta_{\stau}\,\stau_L+\sin\theta_{\stau}\,\stau_R $ and
write the Lagrangian as
\begin{align}
 {\cal L}_{\stau G \tau} + 
 {\cal L}_{\stau G' \tau} 
 &= \frac{m^2_{\stau} - m^2_\tau}{F}
 \left( \cos\theta_{\stau}\, G \tau_L + \sin\theta_{\stau}\, G^\dagger \tau_R\right) \stau^* \nn \\
 &\quad+ 
 \frac{m^2_{\stau} - m^2_\tau}{F}\left(K_{\tau_L} \cos\theta_{\stau}\, G' \tau_L + K_{\tau_R} \sin\theta_{\stau}\, G'^\dagger \tau_R\right) \stau^*
+ \mathrm{ h.c.}, \label{staucoupling}
\end{align}
where we have set $P_{L/R} \tau = \tau_{L/R}$. Once again, we treat the coefficients $K_{\tau_L}$ and $K_{\tau_R}$ as free parameters.

\section{The case of the neutralino LOSP}

We begin our analysis by considering the case where the neutralino $\chi$ is the LOSP. For this we use the Lagrangians (\ref{chiphoton}), (\ref{atal}) and (\ref{ktkl}).

\subsection{Decays of the neutralino}

The partial widths for neutralino decay into a photon and a
(pseudo)-goldstino are:
\begin{align}
\Gamma(\chi \to \gamma G) &= \frac{a_\gamma^2 m_\chi^5}{16\pi F^2},
\\
\Gamma(\chi \to \gamma G') &= \frac{K_{\gamma}^2  a_\gamma^2 m_\chi^5}{16\pi F^2}
\left(1-\frac{m_{G'}^2}{m_\chi^2}\right)^3.
\end{align}
We could also write $F=\sqrt{3}m_{3/2}M_p$ in the denominators above,
with the mass of the gravitino (i.e.~the true goldstino) $m_{3/2}$ and
the reduced Planck mass $M_p=2.43 \times 10^{18}$ GeV.

For a rather massive pseudo-goldstino, the factor between
parenthesis can be significantly smaller than 1, though it is always
of ${\cal O}(1)$ barring any fine tuning of $m_{G'}$ against $m_\chi$.
Hence the branching ratios can be of the same order if
$K_{\gamma}={\cal O}(1)$, or we can have a neutralino decaying almost
exclusively to the pseudo-goldstino if $K_{\gamma}\gg 1$. For instance we
can have $K_{\gamma} \sim 100$ and then the decay rate of the neutralino is
going to be $10^4$ times larger with respect to a single sector
scenario.

In order for the neutralino to decay inside the detector, its total
width cannot be too small. A rough order of magnitude of the bound is
$\Gamma_{\rm tot}\gtrsim 10^{-16}$ GeV. For a single sector scenario and
$m_\chi \sim 200$ GeV, it would translate to $\sqrt{F}\lesssim 10^3$~TeV (see e.g. \cite{Martin:1997ns}),
but in our case the constraint is more flexible due to the
presence of the $K_{\gamma}$ factor.

We now list the partial widths for the decay of the neutralino into a
$Z$ boson and a goldstino or a pseudo-goldstino. For the goldstino,
using either (\ref{derz}) or (\ref{atal}), one obtains the classic
result~\cite{Ambrosanio:1996jn}
\begin{align}
\Gamma(\chi \to Z G)  =  \frac{(2a_{Z_{T}}^2+a_{Z_{L}}^2)m_\chi^5}{32\pi
  F^2}\left(1-\frac{m_Z^2}{m_\chi^2}\right)^4.
\end{align}
For the pseudo-goldstino, the decay rate is given by
\begin{align}
 \Gamma(\chi\to ZG')=\frac{\beta}{16\pi m_\chi}|{\cal M}|^2,
\end{align}
where
$\beta\equiv \beta(\frac{m_Z^2}{m_{\chi}^2},\frac{m_{G'}^2}{m_{\chi}^2})$
with
\begin{align}
 \beta(a,b)=(1+a^2+b^2-2a-2b-2ab)^{1/2}  \label{phasespace}
\end{align}
is the usual phase space factor and the spin summed and averaged amplitude squared is
\begin{align}
|{\cal M}|^2 = \frac{1}{2F^2} & \left\{
(m_\chi^2-m_Z^2)^3(2K_{T}^2
+ K_{L}^2) + 6m_{G'} m_Z (m_\chi^2-m_Z^2)^2K_{T} K_{L}
\right. \nn \\
&  \quad + m_{G'}^2 (m_\chi^2-m_Z^2)\left[ (-4m_\chi^2 - m_Z^2)
  K_{T}^2 +(-2 m_\chi^2 + m_Z^2) K_{L}^2 -6 m_\chi m_Z
  K_{T} K_{L} \right]  \nn\\
& \quad -6 m_{G'}^3 m_Z \left[m_\chi m_Z (K_{T}^2+
  K_{L}^2) +(m_\chi^2 + m_Z^2) K_{T} K_{L}
  \right] \nn \\
&  \quad \left.+ m_{G'}^4 \left[ (2m_\chi^2 + m_Z^2)K_{T}^2+
(m_\chi^2 + 2 m_Z^2)K_{L}^2 +6 m_\chi m_Z K_{T} K_{L}\right]\right\},
\end{align}
where $K_T\equiv K_{Z_T}a_{Z_{T}}$ and $K_L\equiv K_{Z_L}a_{Z_{L}}$.
The decay channel is open only when $m_{G'}<m_{\chi}-m_{Z}$.

We note that, upon setting $K_{\gamma}=K_{Z_T}=K_{Z_L}=1$, the decay
rates to a pseudo-goldstino slightly differ from those to a massive
spin-3/2 gravitino,
as detailed for instance in \cite{hep-ph/0404231}.

\begin{figure}
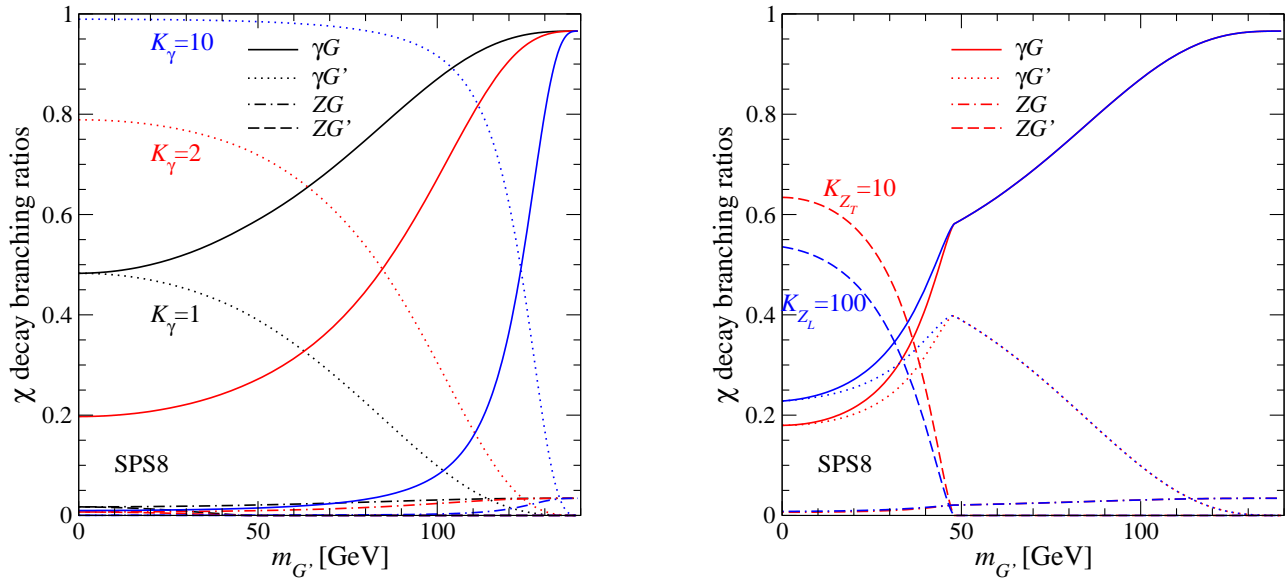

\centering
\includegraphics[width=.45\textwidth,clip]{n1decay_ka.eps}
\hfill
\includegraphics[width=.45\textwidth,clip]{n1decay_kz.eps}
\caption{\small Branching ratios for the lightest neutralino decay into
  a (pseudo)-goldstino and a photon or $Z$ at the SPS8 benchmark point. On the left the coupling of the
  pseudo-goldstino to the photon is gradually enhanced, while on the
  right the case of enhanced coupling of the
  pseudo-goldstino to the $Z$ boson is shown.
  \label{n1decay}}
\end{figure}

In Fig.~\ref{n1decay} we plot the branching ratios for the lightest
neutralino decay for varying pseudo-goldstino mass and for various values of
$K_{\gamma}$, $K_{Z_T}$ and $K_{Z_L}$.
The remaining relevant SUSY parameters are taken at the SPS8 benchmark point as
calculated by SOFTSUSY~\cite{Allanach:2001kg}. Namely, we have
$m_G\,(=m_{3/2})=4.74$ eV, $m_\chi=139.2$ GeV, $N_{11}= 0.99$,
$N_{12}=-0.031$, $N_{13}= 0.124$, $N_{14}= -0.048$ and $\tan\beta= 14.5$.
In general, the partial widths for decay into $Z$ bosons are very much
suppressed with respect to the ones for decay into photons, except
when $K_{Z_T}$ and/or $K_{Z_L}$ are the only large
factors. In the following we will not assume this.
Also, the partial width for decay into higgses is negligible at the
SPS8 point.
See \cite{Thaler:2011me} for a different set up where this is not the case.

It is obvious that the neutralino total width will always be exceedingly
small, e.g.
$\Gamma_{\rm tot}\sim 10^{-12}$ GeV for $K_{\gamma}=K_{Z_T}=K_{Z_L}=1$,
compared to $m_\chi$, so that we can
safely place ourselves in the narrow width approximation (NWA) in all
processes of interest with an intermediate neutralino.
In the next sub-sections we will  discuss observable signatures
involving the production and decay of neutralinos.
Thus, the (differential) cross sections will be
proportional to the square of the amplitudes for production of
neutralinos,
and to the branching ratios for their decay.

\subsection{Goldstini and single photon production in $e^+e^-$ collisions}\label{sec:monoa}

We now perform an analytic computation with the purpose of
highlighting the differences with respect to the single sector
case and the role played by the extra parameters $K_{\lambda}$ and $K_{\phi}$,
characterizing the pseudo-goldstino couplings.
For simplicity, in this subsection only, we stick to the case where the neutralino is a pure
photino, i.e. $a_{\gamma}=1$ and $a_{Z_T}=a_{Z_L}=0$.

\begin{figure}[b]
\centering
\includegraphics[width=\textwidth,clip]{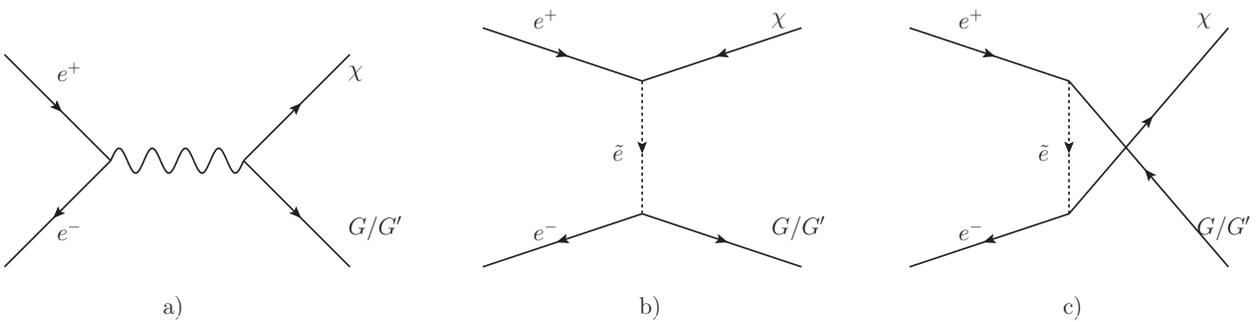}
\caption{\small All diagrams contributing to  $e^+e^-\to \chi G/G'$:
  a) $s$-channel, b) $t$-channel and c) $u$-channel.
\label{allGchi}}
\end{figure}

There are three kinds of diagrams contributing to this process, as in
Fig.~\ref{allGchi}. In the $s$-channel, the intermediate particle is
a photon, since the neutralino is pure photino.
Note also that there is another
$s$-channel diagram with the outgoing arrows reversed. (We are using the two-component notation of~\cite{Dreiner:2008tw}.)
In the $t$- and $u$-channels, the intermediate particle is either of
the right- and left-handed selectrons. We assume
here that the factor
$K_{\phi}$
is the same for both
selectrons.

The couplings of the (pseudo)-goldstino have been reviewed in section~\ref{formal}
and for a pure photino neutralino they are (see (\ref{Glambdacoupl}) and (\ref{Gphicoupl})) 
\begin{align}
\mathcal{L}_{G'} \supset
K_{\lambda} \frac{ m_\chi}{2\sqrt{6} M_p m_{3/2}} \lambda \sigma^\mu
\bar\sigma^\nu G'F_{\mu\nu}+
K_\phi\frac{m^2_{\tilde e}}{\sqrt{3}M_p m_{3/2}}G'\psi \phi^* ,
\end{align}
where $m_{3/2}=F/\sqrt{3}M_{p}$ is the gravitino mass,
$m_{\tilde e}$ is the right and left selectron mass
(we have neglected the mass
of the electron with respect to the mass of the selectrons) and
finally $m_\chi\equiv m_\lambda$.
The other couplings needed for the computation can be
gathered from e.g. \cite{Dreiner:2008tw}.

Note that in the amplitude squared
all diagrams actually interfere due to
the mass of the pseudo-goldstino.
The computation is thus
more involved than the one for a massless goldstino
\cite{Fayet:1986zc,Zichichi}.
The differential cross section
for production of a photino and a pseudo-goldstino reads
\begin{align}
\label{sigmadiff}
  d\sigma_{e^+ e^- \to \chi G'}
 =\frac{1}{2 s}|{\cal M}|^2\,d\Phi_2
\end{align}
with the spin summed and averaged amplitude squared
\begin{align}
 |{\cal M}|^2&= \frac{e^2}{3M_p^2 m_{3/2}^2}\left\{
\frac{K_{\lambda}^2 m_{\chi}^2}{s}
\left( 2 t u - m_{\chi}^2 (t+u)+2 m_{G'} m_{\chi} s + m_{G'}^2 (2
m_{\chi}^2-t-u)\right)\right.  \nonumber\\
& \quad  +   K_{\phi}^2 m_{\tilde e}^4
\left(\frac{(t-m_{G'}^2)(t-m_{\chi}^2)}{(t-m_{\tilde e}^2)^2}+
\frac{(u-m_{G'}^2)(u-m_{\chi}^2)}{(u-m_{\tilde e}^2)^2}
+\frac{2 m_{G'} m_{\chi} s}{(t-m_{\tilde e}^2)(u-m_{\tilde e}^2)}\right)\nonumber \\
& \quad  - \left.
2 K_{\lambda}K_{\phi}m_{\chi}m_{\tilde e}^2
\left(\frac{m_{\chi}(t-m_{G'}^2)+m_{G'}(t-m_{\chi}^2)}{t-m_{\tilde e}^2}
     +\frac{m_{\chi}(u-m_{G'}^2)+m_{G'}(u-m_{\chi}^2)}{u-m_{\tilde e}^2}\right)
\right\},
\end{align}
where $e$ is the electromagnetic coupling constant. We refer to
e.g.~\cite{Nakamura:2010zzi} concerning kinematics.
One can check that in the limit $m_{G'} \to 0$ and $K_{\lambda}=K_{\phi}=1$ this reproduces
the result reported in \cite{Zichichi}.
Plugging $|{\cal M}|^2$ in (\ref{sigmadiff}) and integrating over
$\cos\theta$ we get the total cross section.

It is interesting to note that the above amplitude is slightly
different depending on the relative sign of the respective (real) Majorana
masses of the neutralino LOSP and the pseudo-goldstino. This relative
sign is model dependent.

It should be stressed that the cross section scales with
$K_{\lambda,\phi}^2/m_{3/2}^2$.
In the spirit of~\cite{Zichichi,Dicus:1996ua,Brignole:1997sk,Achard:2003tx,Abdallah:2003np},
LEP bounds on such cross sections
can be translated into upper bounds on $K/m_{3/2}$, or alternatively
on $K$ at a given value of $m_{3/2}$. (Here $K$ is for simplicity a
common value for $K_\lambda$ and $K_\phi$.) Roughly, we get
$K<10^{4-5} (m_{3/2}/\mbox{eV})$, which allows us some elbow room.
Note that in the case of the single true goldstino production, i.e. with
$K_{\lambda}=K_{\phi}=1$,
the cross section is very small, unless the gravitino mass is of order
$10^{-5}-10^{-4}$ eV \cite{Zichichi}.
In the case of the pseudo-goldstino, instead, the cross section can be enhanced by the
couplings $K_{\lambda}$ and $K_{\phi}$
while keeping the gravitino mass to standard values for gauge mediation
scenarios, i.e. $m_{3/2}\sim$~eV.

\begin{figure}
\centering
\includegraphics[width=.475\textwidth,clip]{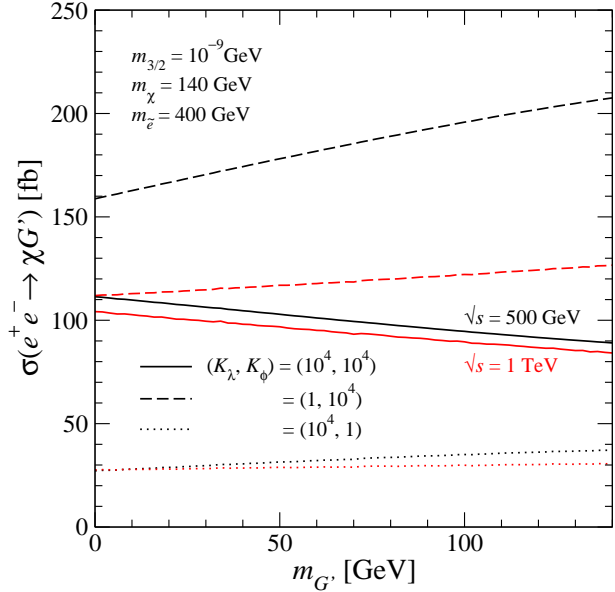}
\caption{\small Total cross section of $e^+e^-\to\chi G'$ at
 $\sqrt{s}=500$~GeV (black lines) and 1~TeV (red lines) as a
  function of the pseudo-goldstino mass, for various values of
 $K_{\lambda}$ and $K_{\phi}$.}
\label{xsec_mpgld}
\end{figure}

To exemplify the physics of this process, and in particular
the dependence on the pseudo-goldstino parameters, in
Fig.~\ref{xsec_mpgld} we plot the total cross section as a function of
the pseudo-goldstino mass for some values of the parameters
$K_{\lambda}$ and $K_{\phi}$.
Here we consider a photino LOSP and take the
masses as
$m_{3/2}=10^{-9}$~GeV, $m_{\chi}=140$~GeV and
$m_{\tilde e_R}=m_{\tilde e_L}=400$~GeV.

There is a destructive interference between the diagrams and thus the
cross section for large $K_\phi$ turns out to be greater than the
cross-section when both $K_\lambda$ and $K_\phi$ are large.
We notice that rather large values of $K_\lambda$ and $K_\phi$ are required
to obtain the cross section around ${\cal O}(10^{2-3})$~fb with the eV
order gravitino mass, while such large values are not favored by the
stability of the SUSY breaking vacuum as mentioned before.

One can also easily see that in these cases
the emitted photons can be significantly softer than in usual gauge
mediation scenario, since the pseudo-goldstino has a non negligible
mass.
Moreover,
similar to the discussions in~\cite{Mawatari:2011cu}, the photon energy
distribution can tell us about the masses of the neutralino and
pseudo-goldstino as we will explicitly see in the next section.
On the other hand, different $K$ factors would give different photon
angular distributions.

All the results presented here can be obtained numerically running
MadGraph 5~\cite{Alwall:2011uj}
simulations adapted to the (pseudo)-goldstino scenario (building on \cite{Mawatari:2011jy}), having implemented
the model using FeynRules \cite{Christensen:2008py,Duhr:2011se,Degrande:2011ua}.
This provides also a non trivial test of our software package, which we will use afterwards
to simulate $pp$ collisions.

\subsection{Goldstini and di-photon production in $e^+e^-$ collisions}\label{sec:diphoton}

The production of two neutralinos will lead to a di-photon plus missing
energy signature, which is  evidence for the processes
\begin{align}
e^+e^- \to \gamma\gamma GG , \qquad e^+e^- \to \gamma\gamma GG' ,
\qquad e^+e^- \to \gamma \gamma G'G' .
\end{align}
The total cross section
($\sigma^{\rm LO}_{e^+e^-\to\chi\chi}\sim 177$ fb at $\sqrt{s}=500$~GeV
at SPS8) is
similar to the single sector case,
since the couplings that can be enhanced in the
pseudo-goldstino scenario only appear in the decay of the neutralinos.
However the photon spectrum, and in particular the
edges of the energy distribution, is sensitive to
the mass of $G'$, both if the branching ratios
are comparable or if the decay to $G'$ is favoured.

\begin{figure}
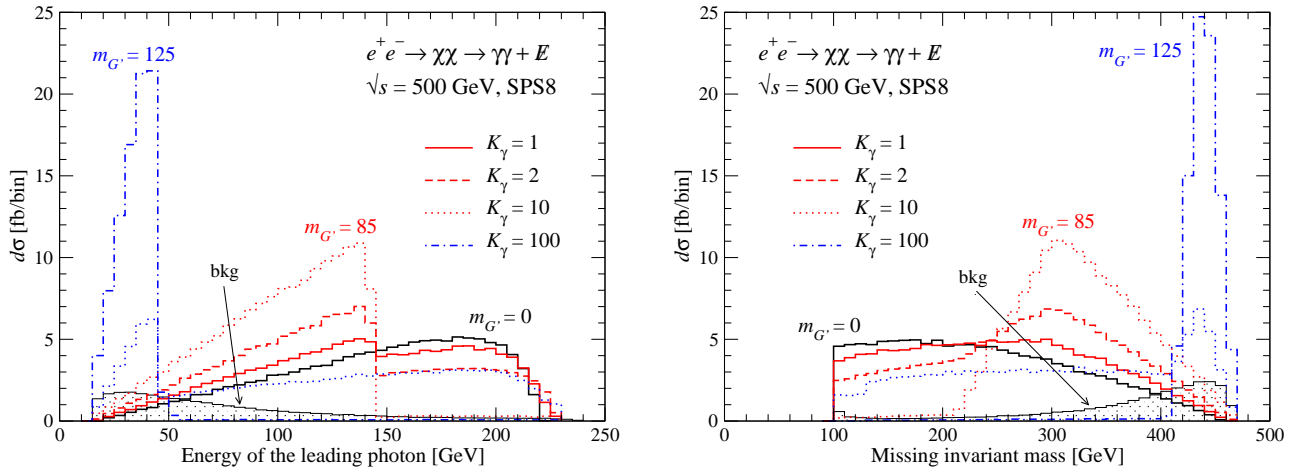

\centering
\includegraphics[width=.48\textwidth,clip]{diphoton_e.eps}
\hfill
\includegraphics[width=.48\textwidth,clip]{diphoton_minv.eps}
\caption{\small
 The distributions of the leading photon energy (left) and of the missing
  invariant mass (right) for
  $e^+e^-\to\chi\chi\to\gamma\gamma+\misse$ at $\sqrt{s}=500$~GeV, where the pseudo-goldstino mass is fixed at 85
 GeV (red) and 125 GeV (blue). The $m_{G'}=0$ case (black) is also shown as the
 reference point as well as the SM background.
  }
\label{ephoton}
\end{figure}

In Fig.~\ref{ephoton},
the distributions of the leading photon energy (left) and
of the missing invariant mass (right) for
  $e^+e^-\to\chi\chi\to\gamma\gamma+\misse$ at $\sqrt{s}=500$~GeV
 are shown.
To obtain both plots,
we applied a cut on the
energy and the rapidity of the photons,
$E_\gamma > 15$ GeV and $|\eta_\gamma|<2$, as the minimal cuts for the
detection of photons.
In addition, we imposed the invisible invariant mass cut
$M_\text{inv}>100$~GeV to remove the SM $(Z\to\nu\bar\nu)\gamma\gamma$ background.
The remaining background comes from the $t$-channel $W$-exchange
process, and this can be reduced by using the polarized $e^{\pm}$
beams.

Besides the reference point $m_{G'}=0$ with $K_\gamma=1$ (for which we
have essentially two
indistinguishable copies of a light goldstino), we take two different
pseudo-goldstino masses, 85~GeV and 125~GeV, with different couplings as
in the following table:
\begin{center}
\begin{tabular}{llll}
 1a. & $m_{G'}=85$ GeV & with $K_{\gamma}=1$ &
       [$B(\chi\to\gamma G')\sim 0.2$] \\
 1b. & $m_{G'}=85$ GeV & with $K_{\gamma}=2$ &
       [$B(\chi\to\gamma G')\sim 0.5$] \\
 1c. & $m_{G'}=85$ GeV & with $K_{\gamma}=10$ &
       [$B(\chi\to\gamma G')\sim 1$] \\
 2a. & $m_{G'}=125$ GeV & with $K_{\gamma}=10$ &
       [$B(\chi\to\gamma G')\sim 0.5$] \\
 2b. & $m_{G'}=125$ GeV & with $K_{\gamma}=100$ &
       [$B(\chi\to\gamma G')\sim 1$]
\end{tabular}
\end{center}
Here, we keep $K_{Z_T}=K_{Z_L}=1$, and hence
the decay modes into $Z$ are negligible;
see also Fig.~\ref{n1decay} for the branching ratios.
We have also run the simulations for $m_{G'}=10$ GeV with
$K_\gamma= 1,2$ and 10, for which the branching ratio is respectively
$0.5$, $0.8$ and $1$, but we found distributions essentially overlapping with the massless one.

The edges of the energy distributions allow to determine both the mass of
the neutralino LOSP and of the pseudo-goldstino. A simple
generalization of the massless goldstino case discussed in
\cite{Ambrosanio:1996jn} gives the following expression for the
minimal and maximal energy of each emitted photon:
\begin{align}
E^{\mathrm{max},\mathrm{min}}_\gamma &=
\frac{\sqrt{s}}{4}\left(1-\frac{m_{G'}^2}{m_\chi^2}\right)
 \left(1\pm\sqrt{1-\frac{4m_\chi^2}{s}}\right).
 \label{edges}
\end{align}
In this case, it is difficult to determine the minimal edges due to the
detector cut. On the other hand, the $E_{\gamma}^{\rm max}$ is rather
clear although the edge of the high energy region is smeared by the
missing invariant mass cut.
It is interesting to note that, unless the branching ratio is not close
to unity, we can find the two $E_{\gamma}^{\rm max}$ edges with
$m_{G'}\ne 0$ and $m_{G'}=0$, which
can determine both $m_\chi$ and $m_{G'}$.
Moreover, we can also determine the branching ratio from the shape of
the distributions, i.e. the information on the coupling.

\subsection{Goldstini production in $pp$ collisions}
\label{sec:pp_chichi}

We now turn to consider the processes which are relevant to the LHC.

Similar to the process $e^+e^-\to \chi G'$ in Sec.~\ref{sec:monoa}, the
cross section of $pp\to\chi G'$
is proportional to $K^2/m_{3/2}^2$ and could be enhanced by the factor
$K$.
However, rather large $K$ values of ${\cal O}(10^{4-5})$ are needed to obtain
a visible cross section for an eV-order gravitino, leading to a
bound for the $K$ values
by the $\gamma+\misset$ events at the Tevatron~\cite{Acosta:2002eq}
similar to the LEP bound
discussed above.
We note that
at the parton level
the amplitudes are
the same as the ones studied in the previous section
after replacing the incoming electrons with the quarks.\footnote{Note that only in this process, and the following one, we would need to specify some squark masses. These can also be easily extracted from SPS8, however as we will argue the most interesting process that we study in more details does not involve intermediate coloured superpartners.}

The clean $\gamma\gamma+\misset$ signal is given by the neutralino LOSP
pair production.
However, the cross section is too small
($\sigma^{\rm LO}_{pp\to\chi\chi}\sim 0.3\,(1.2)$ fb at $\sqrt{s}=7\,(14)$~TeV
at SPS8) to be significant over the SM background.%
\footnote{It is well known that the NLO QCD corrections enhance the LO cross
section by a factor of 1.3-1.4 for $\sqrt{s}=14$ TeV~\cite{Beenakker:1999xh}.
It is also interesting to note that $gg$ collisions can give a certain
contribution to the neutralino pair production through one
loop~\cite{Yi:2000dt}.
The corresponding amplitudes are suppressed by the loop factor, while
they are enhanced because of the larger gluon PDF, and also because
all (s)quarks can run in the loop.
}
It is worth to emphasize again that the emitted photons associated with a
pseudo-goldstino are softer than those with a true goldstino,
which makes it difficult to apply some cuts to enhance the signal over the
background.
In other words, the experimental constraints for the SPS8 point in the
standard minimal GMSB model as well as for
general gauge mediation \cite{Meade:2008wd}
(e.g. \cite{Abel:2009ve,Abel:2010vba,arXiv:0911.4130,arXiv:1103.6083,Kats:2011qh})
could be eased in the multiple goldstini
scenario.

Among the exclusive processes with at most two extra particles in the
final state, the cleanest one is the one where the two photons and
missing energy are accompanied by two leptons.
By far the main contribution to the signal comes from the
pair production of sleptons, which subsequently decay into a lepton and
the neutralino LOSP,
$pp\to\tilde l^+_{R/L}\tilde l^-_{R/L}\to l^+l^-+\gamma\gamma+\misset$ ($l=e,\mu$).
The electron and muon pairs give the same contributions due to the
degeneracies between the first two slepton families.
The Standard Model background is completely negligible compared to the signal of new physics.

\begin{figure}
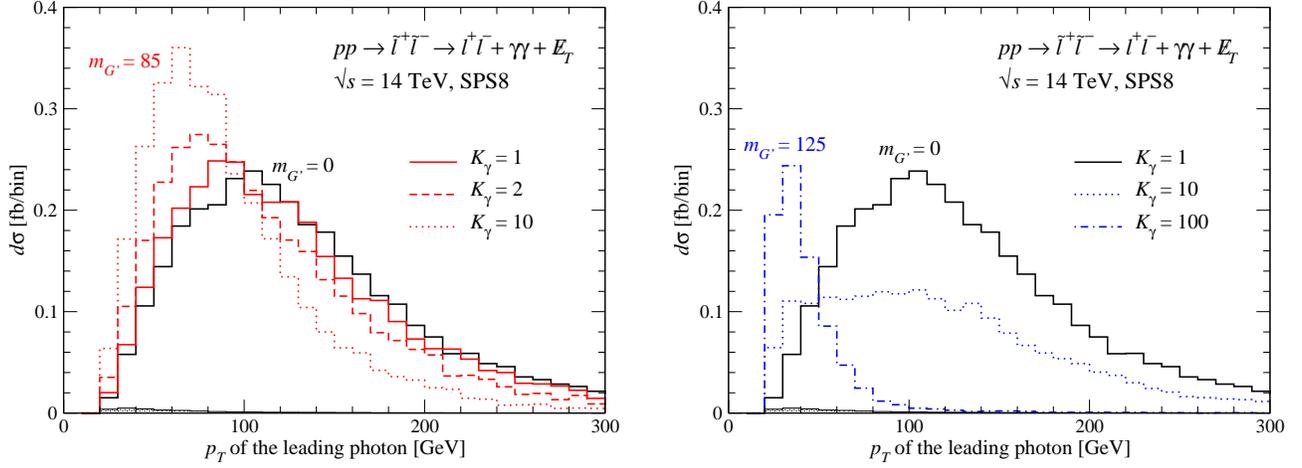

\centering
\includegraphics[width=.48\textwidth,clip]{LHC14_pt_m85.eps}
\hfill
\includegraphics[width=.48\textwidth,clip]{LHC14_pt_m125.eps}
\caption{\small The $p_T$ distributions of the leading photon for
 $pp\to\tilde l^+\tilde l^-\to l^+l^-+\gamma\gamma+\misset$ at
 $\sqrt{s}=14$~TeV for $m_{G'}=85$~GeV (left) and 125~GeV (right).
 The $m_{G'}=0$ case is also shown as a reference with black lines.
 }
\label{pt1LHC14}
\end{figure}

\begin{figure}
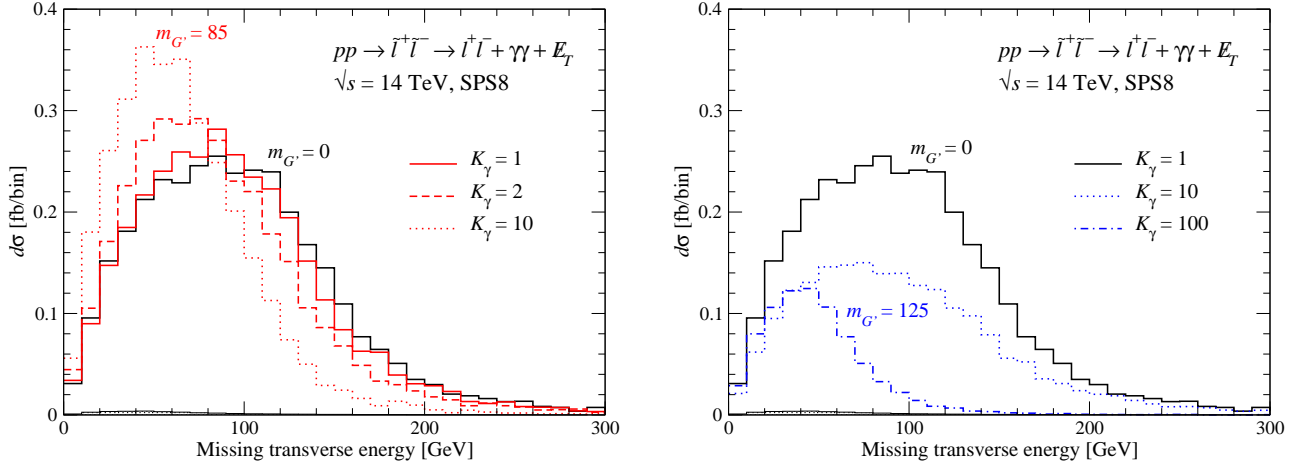

\centering
\includegraphics[width=.48\textwidth,clip]{LHC14_met_m85.eps}
\hfill
\includegraphics[width=.48\textwidth,clip]{LHC14_met_m125.eps}
\caption{\small The same as Fig.~\ref{pt1LHC14}, but the missing
 transverse energy distribution.
 }
\label{metLHC14}
\end{figure}

The signal cross
section is 1.1 (3.3) fb at $\sqrt{s}=7$ (14) TeV at SPS8, where
the slepton masses are $m_{\tilde l_{R/L}}=180.2/358.2$ GeV and
we employ the CTEQ6L1 PDFs~\cite{Pumplin:2002vw}
with the factorization scale chosen as
$\mu=(m_{\tilde l_R}+m_{\tilde l_L})/2$.%
\footnote{The NLO cross section is about 1.35 times larger than the LO
one at $\sqrt{s}=14$~TeV~\cite{Baer:1997nh,Beenakker:1999xh}. The gluon
fusion contribution to the slepton pair productions has been also
studied in~\cite{delAguila:1990yw,Borzumati:2009zx}.\label{sleptonnlo}} 
Here, the leptons and
photons are required to have $p_T>20$~GeV, $|\eta|<2.5$, and
$R_{ll,\gamma\gamma}>0.4$,
where $p_T$ and $\eta$ are the transverse momentum and the
pseudorapidity of a final-state particle, respectively, and
$R_{ij}$ describes the separation of the two particles in the plane
of the pseudorapidity and the azimuthal angle.

Because of the peculiarities of hadronic collisions, raising the
center-of-mass energy from $7$~TeV to $14$~TeV increases the total
cross section but does not alter significantly the shape of the
distributions.
Therefore, we present the results only for the 14 TeV LHC here.
In Figs.~\ref{pt1LHC14} and \ref{metLHC14}, the distributions of
the $p_T$ of the leading photon and of the
transverse missing energy are shown, respectively, where
the same benchmark points are taken as in Sec.~\ref{sec:diphoton}.
The SM background is invisibly small.
The two plotted variables show different
shapes with respect to the standard goldstino scenario ($m_{G'}=0$)
depending on the mass and the coupling of the pseudo-goldstino.
This fact could in principle be used to extract the
masses and the couplings by using techniques explained
in~\cite{hep-ph/9906349,arXiv:0802.2879,arXiv:1004.2732,arXiv:1109.3471}.
We note that in the case of $m_{G'} = 125$~GeV the signal cross section is
largely reduced by the experimental cuts, especially for large
$K_{\gamma}$.
This is due to the fact that, for more massive pseudo-goldstino,
the emitted photons will be softer, and hence more excluded by the
kinematical cuts.

Before closing the section,
we note that the production cross sections of the colored SUSY particles
at SPS8 are small due to their large masses in the TeV range.
In general, although we have presented the exclusive signals without jets here,
the inclusive search is also interesting and will be reported elsewhere.

\section{The case of the stau LOSP}

We now move on to the case where the $\stau$ is the LOSP, and use the
Lagrangian (\ref{staucoupling}) to study it. 
The effect of the coefficients $K_{\tau_L}$ and $K_{\tau_R}$ is that
of enhancing the $G'$ production.

\subsection{Decays of the stau}

\begin{figure}
\centering
\includegraphics[width=.45\textwidth,clip]{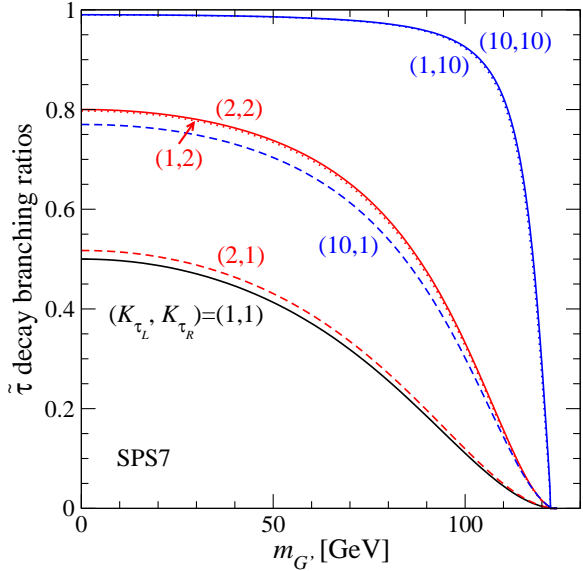}
\caption{\small Branching rations of $\stau\to\tau G'$ at the SPS7
 benchmark point, for various values of $K_{\tau_L}$ and $K_{\tau_R}$. 
 }
\label{staudecayplots}
\end{figure}

The decay amplitude squared for the process $\stau\to\tau G'$
can be easily computed in terms of the coefficients in the Lagrangian
yielding 
\begin{align}
 |{\cal M}|^2 =\left(\frac{m^2_{\stau} - m^2_\tau}{F}\right)^2\big(&
   ( K^2_{\tau_L}\cos^2\theta_{\stau} 
    +K^2_{\tau_R}\sin^2\theta_{\stau})(m^2_{\stau} - m^2_{G'} - m^2_{\tau}) \nn \\
    & + 4 K_{\tau_L} K_{\tau_R}\sin\theta_{\stau}\cos\theta_{\stau}\,
         m_{G'}m_{\tau} \big).
\end{align}
The width can be written as
\begin{align}
 \Gamma(\stau\to\tau G') = \frac{\beta}{16 \pi m_{\stau}} |{\cal M}|^2
\end{align}
with
$\beta\equiv \beta(\frac{m_{\tau}^2}{m_{\stau}^2},\frac{m_{G'}^2}{m_{\stau}^2})$
as in (\ref{phasespace}).
Notice that the width depends not only on the magnitude but also on the ratio of the coefficients $K_{\tau_L}$ and $K_{\tau_R}$.
The well known expression for the true goldstino case is obtained from
the above by simply setting $K_{\tau_L}=K_{\tau_R} = 1$ and 
$m_{G'}=0$.  

For illustration we plot the branching ratios of the decay 
$\stau\to\tau G'$, choosing the stau mass and mixing angle given by the
SPS7 benchmark point, namely $m_{\stau} = 124.0$ GeV and $\cos \theta_{\stau}=0.154$. The smallness of  $\cos \theta_{\stau}$ at the SPS7 point makes the branching ratio largely independent from $K_{\tau_L}$ but this would of course change for a different mixing angle.

The total width of the stau LOSP at SPS7, where $m_G(=m_{3/2})=0.76$~eV,
is of the order of $10^{-10}$~GeV for $K_{\tau_L}=K_{\tau_R}=1$, and hence
the stau will decay promptly into the detector. Scenarios with long
lived staus could also be envisioned, with even the possibility of
stopping them and then measuring their decays. In that case the
branching ratio and the mass of the invisible particle(s) should be more
directly accessible. We will not delve further on this scenario.  

\subsection{Goldstini and di-tau production}

We now turn to stau production at colliders and subsequent decay into
taus and (pseudo)-goldstini. 
Though we could first consider stau pair production at
$e^+e^-$ colliders, we will omit this rather straightforward
exercise. In the neutralino LOSP scenario, the benefit of this set up
was the ability to infer the neutralino and pseudo-goldstino masses from
the energy distribution of the emitted photons. In the present case
the same considerations are practically harder to achieve because of the
difficulties inherent in tau reconstruction. We thus proceed to consider
directly the case of hadron colliders. 

\begin{figure}
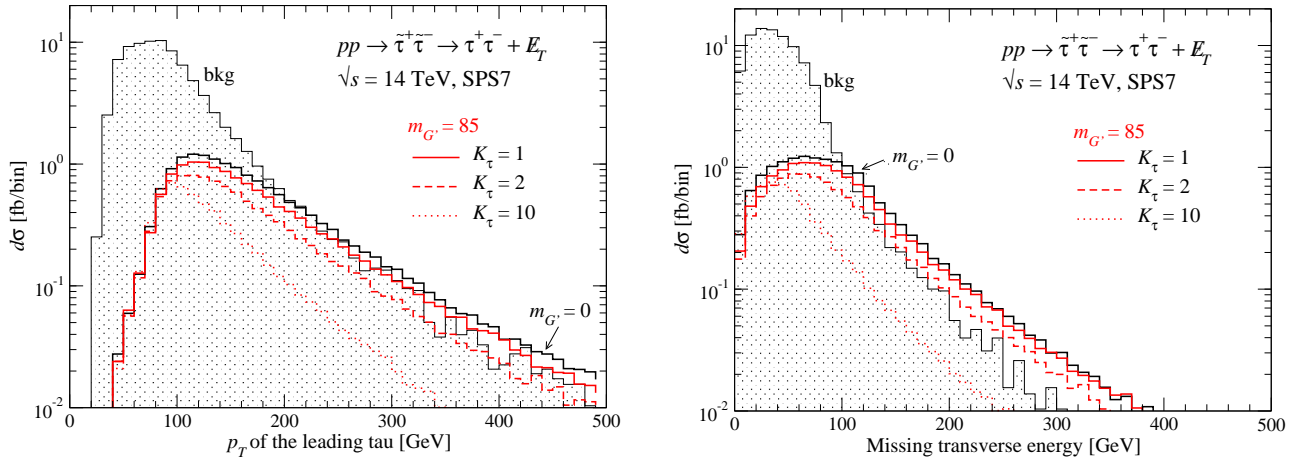

\centering
\includegraphics[width=.48\textwidth,clip]{LHC14_pt_m85_stau.eps}
\hfill
\includegraphics[width=.48\textwidth,clip]{LHC14_met_m85_stau.eps}
\caption{\small The distributions of  
$p_T$ of the leading tau (left) and of the missing transverse energy
 (right) for
 $pp\to\stau^+\stau^-\to\tau^+\tau^-+\misset$ at
 $\sqrt{s}=14$~TeV for $m_{G'}=85$~GeV.
 The $m_{G'}=0$ case (black) is also shown as the
 reference point as well as the SM background.
 }
\label{pt1tauLHC14}
\end{figure}

The process we are interested in is
$pp\to\stau^+\stau^-\to\tau^+\tau^-+\misset$.  
The lighter stau-pair production cross section at SPS7 is 41~fb at
$\sqrt{s}=14$~TeV,%
\footnote{See footnote~\ref{sleptonnlo}, and also~\cite{Lindert:2011td}
 for more details on the stau-pair production.}   
while the irreducible SM background $\tau^+\tau^-\nu\bar\nu$, mainly
coming from 
$Z$ and $W$ pair productions, is ten times larger with the minimal
cuts $p_{T_\tau}>20$~GeV, $|\eta_{\tau}|<2.5$, and $R_{\tau\tau}>0.4$.
Since the background is located in the low tau-tau invariant mass
region, we impose the additional cut $M_{\tau\tau}>150$~GeV in the
analyses below.   
Figure~\ref{pt1tauLHC14} shows the distributions of the $p_T$ of the
leading tau and of the missing transverse energy for the following 
benchmark points covering different branching ratios for stau decays
into $G$ or $G'$:  
\begin{center}
\begin{tabular}{llll}
 1a. & $m_{G'}=85$ GeV & with $K_{\tau_L}=K_{\tau_R}=1$ &
       [$B(\stau\to\tau G')\sim 0.2$] \\
 1b. & $m_{G'}=85$ GeV & with $K_{\tau_L}=K_{\tau_R}=2$ &
       [$B(\stau\to\tau G')\sim 0.5$] \\
 1c. & $m_{G'}=85$ GeV & with $K_{\tau_L}=K_{\tau_R}=10$ &
       [$B(\stau\to\tau G')\sim 1$]
\end{tabular}
\end{center}
The reference point $m_{G'}=0$, i.e. the standard goldstino scenario, is
also shown by a black solid line.
As in the neutralino LOSP scenario, a sizable branching ratio of stau
decays to pseudo-goldstini has an impact on the shape of the
distribution, once again making the spectrum softer. 
Another noticeable fact is that the new physics signal becomes dominant
especially for high missing transverse energy only when the stau decays
to true goldstinos are significant. 
Therefore, the pseudo-goldstini scenario makes it harder to achieve
enough significance for its observation. 

\begin{table}
\centering
\begin{tabular}{|l||r|rrr|r|}
\hline
  & $m_{G'}=0$ & & $m_{G'}=85$ & & background \\ 
  & & $K_{\tau}=1$ & $K_{\tau}=2$ & $K_{\tau}=10$ &  \\ \hline
 $\sigma$ [fb] & 15.5 & 13.2 & 10.1 & 5.9 & 86.4 \\
 $s/b$ & 0.18 & 0.15 & 0.12 & 0.07 & \\
 $S$ (100 (30) fb$^{-1}$) & 5.1 (2.8) & 4.4 (2.4) & 3.4
         (1.8) & 2.0 (1.1) & \\ \hline
\end{tabular}
\caption{Cross sections for the signals and the background. 
 The significance $S$ in~\eqref{signif} 
  is calculated with an integrated luminosity of 100
 (30) fb$^{-1}$ and 
 a di-tau detection efficiency $A=0.1$.}
\label{table:etmiss0}
\end{table}
 
In order to quantify the sensitivity of the missing transverse energy cut,
in Tables \ref{table:etmiss0} and \ref{table:etmiss100} we list the
cross sections, signal ($s$) over background ($b$) and significances
($S$) for our benchmarks, without and with a cut on $\misset$ at 100 GeV
respectively.   
Taking into account the branching ratio of the hadronic tau decays,
$B(\tau_{had})=0.648$~\cite{Nakamura:2010zzi}, and
the efficiency of the hadronic tau identification, 
$\epsilon=0.5$~\cite{CMS,ATLAS}, we assume the di-tau detection
efficiency $A=(B(\tau_{had})\times \epsilon)^2\sim0.1$. 
We use the signal significance defined as~\cite{Cowan:2010js}
\begin{align}
 S=\sqrt{2((s+b)\ln(1+s/b)-s)}.
\label{signif}
\end{align}
We observe that for 100 fb$^{-1}$ of integrated luminosity, a
satisfactory significance can be achieved for most of our benchmarks,
especially for $\misset > 100$~GeV.  
However, in the $K_\tau = 10$ case, i.e. in the case of the heavy
pseudo-goldstino with the enhanced coupling, 
the cut on $\misset$ does not improve the significance of the signal.  
 
\begin{table}
\centering
\begin{tabular}{|l||r|rrr|r|}
\hline
  & $m_{G'}=0$ & & $m_{G'}=85$ & & background \\ 
  & & $K_{\tau}=1$ & $K_{\tau}=2$ & $K_{\tau}=10$ &  \\ \hline
 $\sigma$ [fb] & 5.6 & 4.8 & 3.4 & 0.9 & 3.5 \\
 $s/b$ & 1.59 & 1.36 & 0.95 & 0.25 & \\
  $S$ (100 (30) fb$^{-1}$) & 7.9 (4.3) & 6.9 (3.8) & 5.0
         (2.7) & 1.4 (0.8) & \\ \hline
\end{tabular}
\caption{The same as Table~\ref{table:etmiss0}, 
 but with the additional $\misset>100$~GeV cut.}
\label{table:etmiss100}
\end{table}

Finally we note that, although we took $K_{\tau_L}=K_{\tau_R}$ for
simplicity, tau polarization may be exploited to determine the 
case that the two coefficients are different, 
$K_{\tau_L}\ne K_{\tau_R}$~\cite{Choi:2006mt}.

\section{Conclusions}

In this paper we have analyzed how the
traditional expectations from a low scale gauge mediated model
can be modified by the presence of a massive pseudo-goldstino NLSP in a scenario with multiple-sector SUSY
breaking.

We first considered the case of a gaugino-like neutralino LOSP.
We showed that the decay modes of the LOSP into a photon or $Z$-boson and a pseudo-goldstino can be significant.
We studied in details the goldstini phenomenology in the photon(s) plus missing energy signals
in $e^+e^-$ and $pp$ collisions.
Our aim was to provide clues to interpret prompt photon plus missing energy signals at the LHC.
We found that the resulting photon spectrum is typically softer and with different shapes,
compared to the standard gauge mediation scenario with only one hidden sector.

We then proceeded to consider the case of a stau LOSP. Similarly to the previous case, the stau can significantly decay into a tau and a pseudo-goldstino. We studied a possible tau pair plus missing energy signal at the LHC, again finding that the pseudo-goldstino scenario leads to an altered shape in the tau energy distributions. The signature however seems more promising in the neutralino LOSP scenario, not only because photons have cleaner experimental features than taus, but also because the SM background is virtually absent in the process that we studied.


Although we performed our analyses at the particular benchmark points
(SPS8 for $\chi$ LOSP and SPS7 for $\stau$ LOSP) to make a comparison
with the standard gauge mediation model, our goldstini model can be
easily extended to other benchmark points and LOSP scenarios due to the
implementation into a event generator, MadGraph 5.
We emphasized the dependence of the experimental signatures on the new
parameters, namely the pseudo-goldstino mass and its couplings. 
A rather model independent feature of the pseudo-goldstini scenario is that it provides a scenario which is fully consistent with gauge mediation of supersymmetry breaking, while it softens significantly the energy of the SM products of the LOSP decays, due to the conspicuous mass of the pseudo-goldstino. In this way such a scenario evades most experimental bounds that tend to exclude gauge mediated scenarios. 

We would like to suggest that models like the one
we have presented here, where the missing energy is carried away by two
different particles, one rather massive and the other almost massless, and
with similar couplings but with different strength, deserve to be considered in LHC data analysis.

\subsection*{Acknowledgments}

We would like to thank H.~Haber, Z.~Komargodski, F.~Maltoni,
N.~Seiberg and J.~Thaler for stimulating discussions. We are also
grateful to J.~Alwall, R.~Frederix and O.~Mattelaer for help with MadGraph, to
C.~Duhr and B.~Fuks for help with FeynRules, and to F.~Blekman and
H.~Yokoya for suggestions on the tau detection.
Y.T would like to thank Fabio Maltoni and the members of the CP3,
Universite Catholique de Louvain for their warm hospitality,
where a part of this work was done.

The research of R.A. is supported in part by IISN-Belgium (conventions
4.4511.06, 4.4505.86 and 4.4514.08). R.A. is a Research Associate of the Fonds de la Recherche Scientifique--F.N.R.S. (Belgium).
The research of G.F. is supported in part by the Swedish Research Council (Vetenskapsr{\aa}det)
contract 80409701.
A.M. is a Postdoctoral Researcher of FWO-Vlaanderen.
A.M. is also supported in part by FWO-Vlaanderen through project G.0428.06.
K.D.C. and K.M. are supported by the Concerted Research action
``Supersymmetric Models and their Signatures at the LHC''
of the Vrije Universiteit Brussel.
R.A., K.D.C, A.M. and K.M. are supported in part by the Belgian Federal Science Policy Office
through the Interuniversity Attraction Pole IAP VI/11.
Y.T. was also supported in part by Sokendai short-stay study abroad program.

\end{document}